# Cascade PID Control of an Inverted Pendulum on a Cart System: Simulation and Experimental Analysis

Khalid Mehrab
*Faculty of Computing, Engineering and Science*
*University of South Wales*
Pontypridd, Wales
khalid.mehrab@southwales.ac.uk

Md Zamiul Alam
*Banki Donat Faculty of Mechanical and Safety Engineering*
*Obuda University*
Budapest, Hungary
zamiulalam22@stud.uni-obuda.hu

Shadman Tahmid Haque, Alumnus
*Faculty of Computing, Engineering and Science*
*University of South Wales*
Pontypridd, Wales
shadmantahmid1234@gmail.com

***Abstract*—This study investigates the performance of cascade PID control architecture applied to an inverted pendulum on a cart system through both simulation and experimental implementation. A nonlinear model of the system was developed using Simscape Multibody in Simulink, while a physical prototype was constructed using a DC motor-driven cart, pendulum, rotary encoder, ultrasonic sensor, and an Arduino. The cascade PID control structure consists of an inner loop regulating the pendulum angle and an outer loop controlling the cart position. Simulation results demonstrated effective stabilization of the pendulum and satisfactory position tracking under idealized conditions. Experimental results confirmed successful real-time stabilization but revealed notable differences from simulation, particularly in controller gains, transient behavior, and disturbance response due to sensor noise, unmodeled friction, and implementation constraints. The study also highlights the limitations of cascade PID control in disturbance rejection and large position commands, particularly under limited track length. A comparative analysis using an LQR-based inner loop demonstrated better disturbance rejection and reduced overshoot. The results provide practical insights into the applicability and limitations of cascade PID control of the inverted pendulum system.**

## I. Introduction

The inverted pendulum on a cart is a classic benchmark problem in control theory and robotics, representing a highly non-linear, multivariable, and underactuated mechanical system [1-5]. It is modeled as a Single-Input Multi-Output (SIMO) system in which one control force applied to the cart regulates two outputs: the cart's horizontal position and the pendulum angle [6-8]. The system has two inherent equilibrium states: a stable one where the pendulum hangs downward without requiring any control, and an unstable upper equilibrium where the mass is balanced above the pivot [9, 10]. In the inverted state, the system is inherently unstable and will fall under gravity unless active control is applied to maintain balance [11, 12].

This balancing problem is central to many high-precision technologies and mechatronic applications [13, 14]. The dynamics of the inverted pendulum model complex behavior seen in many real-world applications, including attitude stabilization of launch vehicles during lift-off, missile guidance, gait control in humanoid and legged robots, and balance control in self-balancing vehicles such as the Segway [13-20]. As it involves uncertainty, strong coupling, and non-minimum phase behavior, it serves as an effective experimental platform for evaluating the robustness of different control strategies [9, 15, 21].

To implement a controller, an accurate mathematical model must be derived using Newton's second law to obtain the transfer function and state-space representation [1, 9, 22]. Although experimental validation is preferred in the control community for testing realistic scenarios, many studies rely on MATLAB-Simulink simulations because of the high costs of physical setups [3, 12, 23]. These simulations enable engineers to identify parameters that are difficult to measure and assess the effectiveness of various control strategies in improving stability and response speed before practical implementation [6, 23, 24].

Despite the emergence of advanced algorithms such as Linear Quadratic Regulators (LQR), fuzzy logic, sliding mode control, and neural networks, the Proportional-Integral-Derivative (PID) controller continues to be the most widely used in both industry and academia [12, 13, 23-26]. The PID algorithm is preferred for its simple structure, clear functionality, and user-friendly nature, making it a standard solution for many practical problems [3, 12]. For the inverted pendulum system, the PID control algorithm is composed of three components: the P-term (proportional to error), the I-term (proportional to the integral of error), and the D-term (proportional to the derivative of error) [1].

Since the inverted pendulum is a SIMO system while the standard PID is designed for SISO applications, researchers often use a double-PID or cascade structure [4, 6, 12]. This approach generally uses two control loops: a PID controller for the pendulum angle to stabilize the pole, and another one for the cart to regulate its position [3, 4]. However, due to the strong coupling between the pendulum and cart dynamics, adjusting parameters is often challenging, since adjustments in one controller can affect the performance of the other [3, 27]. Moreover, although PID performs well for small disturbances, it often struggles with the system's inherent nonlinearities and external disturbances, leading researchers to explore more robust and adaptive hierarchical control strategies to enhance reliability in practical situations [14, 25, 27-29].

This study provides insight into a cascade PID control method applied to an inverted pendulum on a cart through both simulation and experimentation, highlighting the practical factors that cause differences between the two approaches. It also examines the impact of high demands on the inner loop controller in systems where a large outer loop overshoot is physically constrained, particularly in terms of achieving system stability.

## II. Non-linear Model Design

The inverted pendulum on a cart is a nonlinear two-dimensional system. The cart moves horizontally, while the pendulum rotates freely in the vertical plane about a pivot mounted on the cart, as illustrated in Fig. 1. The system outputs are the cart position (x) and the pendulum angle (Theta), with the applied force (F) acting as the control input. The upright equilibrium position of the pendulum corresponds to zero degrees.

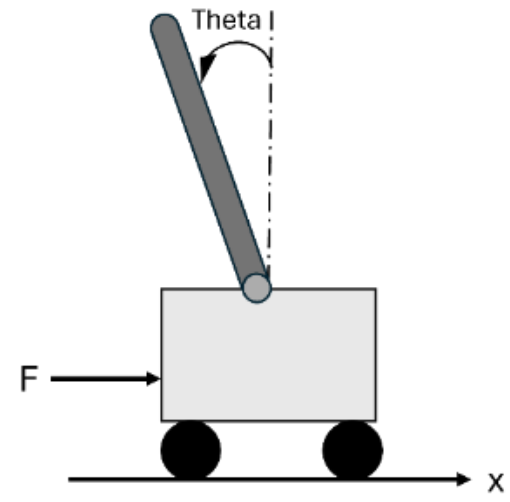

Fig. 1. Inverted pendulum on a cart.

Based on the study of A. N. K. Nasir et al. [30], the parameters of the inverted pendulum system were selected, as listed in Table I.

TABLE I. Dimensions of the Inverted Pendulum System

| Parameter | Value |
|---|---|
| Mass of the cart, M | 0.5672 kg |
| Mass of the pendulum, m | 0.0374 kg |
| Length of the pendulum, l | 0.38 m |
| Moment of inertia of the pendulum, I | $4.5017 \times 10^{-4}$ kg.m$^2$ |

The parameters from Table I were used to develop a non-linear inverted pendulum model in Simulink using Simscape Multibody. The model was designed based on a MATLAB example discussed in [31], as shown in Fig. 2. Simscape Multibody helps to model the physical system without requiring analytical derivation or linearized equations of motion. External disturbances (dF) were applied to the model using the Simulink Signal Editor.

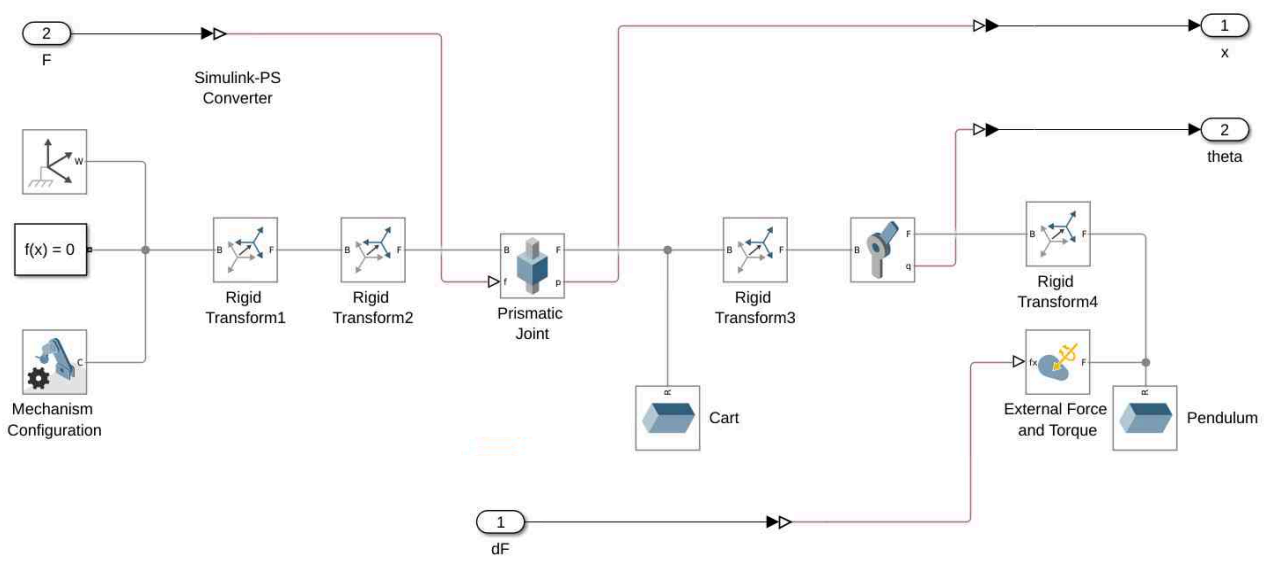

Fig. 2. Simscape multibody model of the inverted pendulum on a cart system.

## III. Experimental Setup

The system was constructed for experimental evaluation and consists of the following components, shown in Fig. 3.

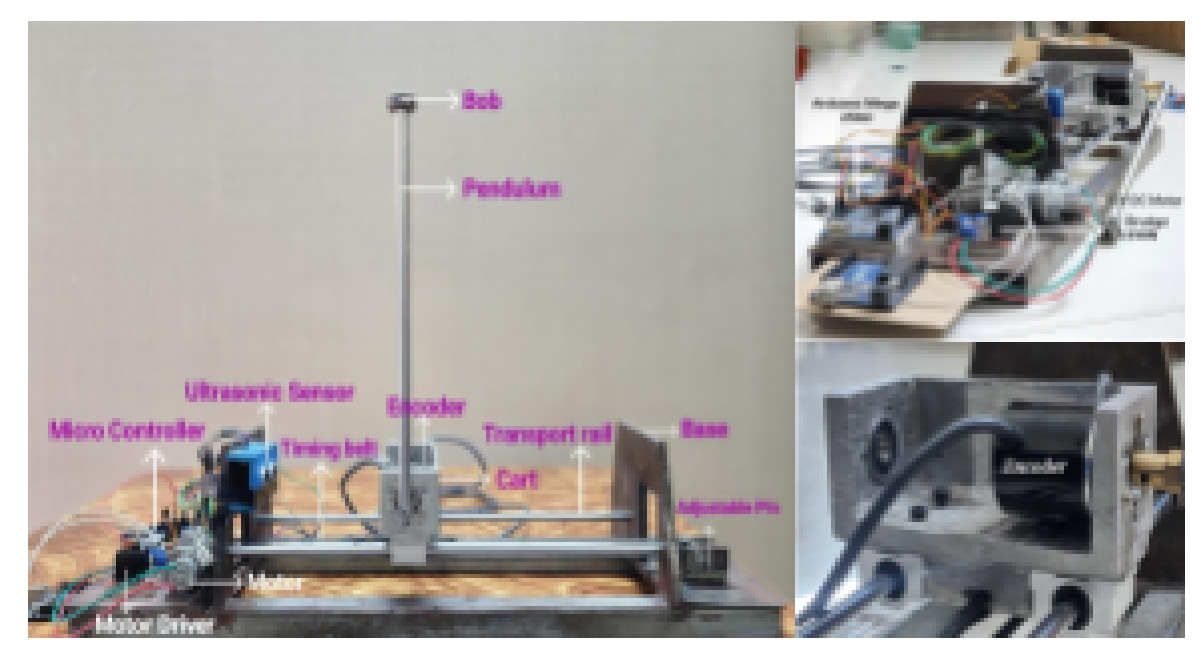

Fig. 3. Experimental setup of the inverted pendulum on a cart system.

Table II lists the key hardware components of the system and their purpose.

TABLE II. Components of the System

| Components | Purpose |
|---|---|
| H-bridge (L298N) | Changes the motor speed and direction. |
| DC geared motor (12V, 800 RPM) | Allows the movement of the cart along the transport rail. |
| Rotary Encoder (Omron E6B2-CWZ6C, 600 ppr) | Determines the angle of the pendulum. |
| Arduino MEGA | Communicates with the hardware to establish the control mechanism. |
| Ultrasonic sensor HC-SR04 | Measures the linear position of the cart. |
| A transport rail | Provides low-friction movement of the cart. |

The rotary encoder produced 2400 pulses per full revolution, with anticlockwise and clockwise rotations defined as positive and negative, respectively. The downward equilibrium position corresponded to −1200 pulses.

## IV. Controller Design

The system was controlled using the Cascade PID approach in both simulation and experiment. This method employs two nested control loops, allowing fast corrections to operate independently of slower movements for a single reference input. The control system was designed in Simulink as illustrated in Fig. 4.

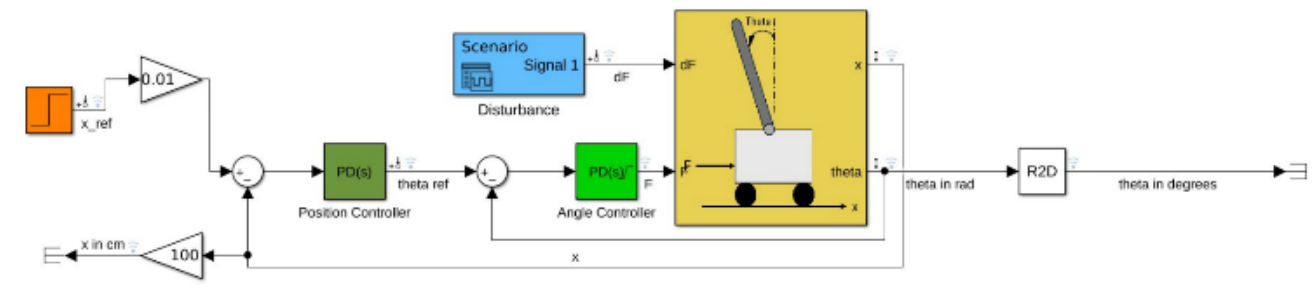

Fig. 4. Designed Cascade PID control system in Simulink.

The inner loop regulates the pendulum angle while the outer loop controls the cart position. The ‘Angle Controller’ was tuned for a higher effective bandwidth than the ‘Position Controller’ to avoid instability or oscillations. The controller's objective was to maintain the pendulum vertically upward when the cart was subjected to a reference input of 10 cm. PID controllers were implemented using standard Simulink PID blocks, with the angle controller output limited to ±12N. This limit was estimated based on the DC motor voltage output. Initial controller gains were determined using the built-in PID Tuner with time-domain settings, followed by manual tuning to achieve minimal (<10%) overshoot for the cart position and <15% angular deviation for the pendulum. The angle controller was tuned to achieve a desired settling time, which was ≤ 3 seconds.

In the experiment, a cascade PID controller was implemented using the Arduino IDE's built-in PID library. A linear setpoint for the cart position was applied to the position controller, which generated the corresponding rotational setpoint for the pendulum, tracked by the angle controller. Controller outputs were scaled from -255 to 255 Pulse Width Modulation (PWM) and remapped to motor commands between 50 and 255 PWM, with the sign indicating the direction of motion. The control signal drove the motor via the motor driver to move the cart along the rail. Both linear and angular positions were measured and compared to their respective setpoints by the controllers.

## V. Results and Analysis

This section presents both simulation and experimental results obtained using the cascade PID control architecture. The analysis evaluates the effectiveness of the control structure by comparing simulated and experimental system behavior.

### *A. Simulation Results*

Fig. 5 illustrates the simulated time response of the pendulum angle. The pendulum angle exhibited a stabilizing response with an absolute overshoot of 10.44°. After the initial overshoot, the angle was quickly corrected by reaching around 2° and moving to equilibrium. This indicates a robust and stable inner loop response. The rise time is moderate, followed by a gradual decay to a steady state within approximately 3 seconds.

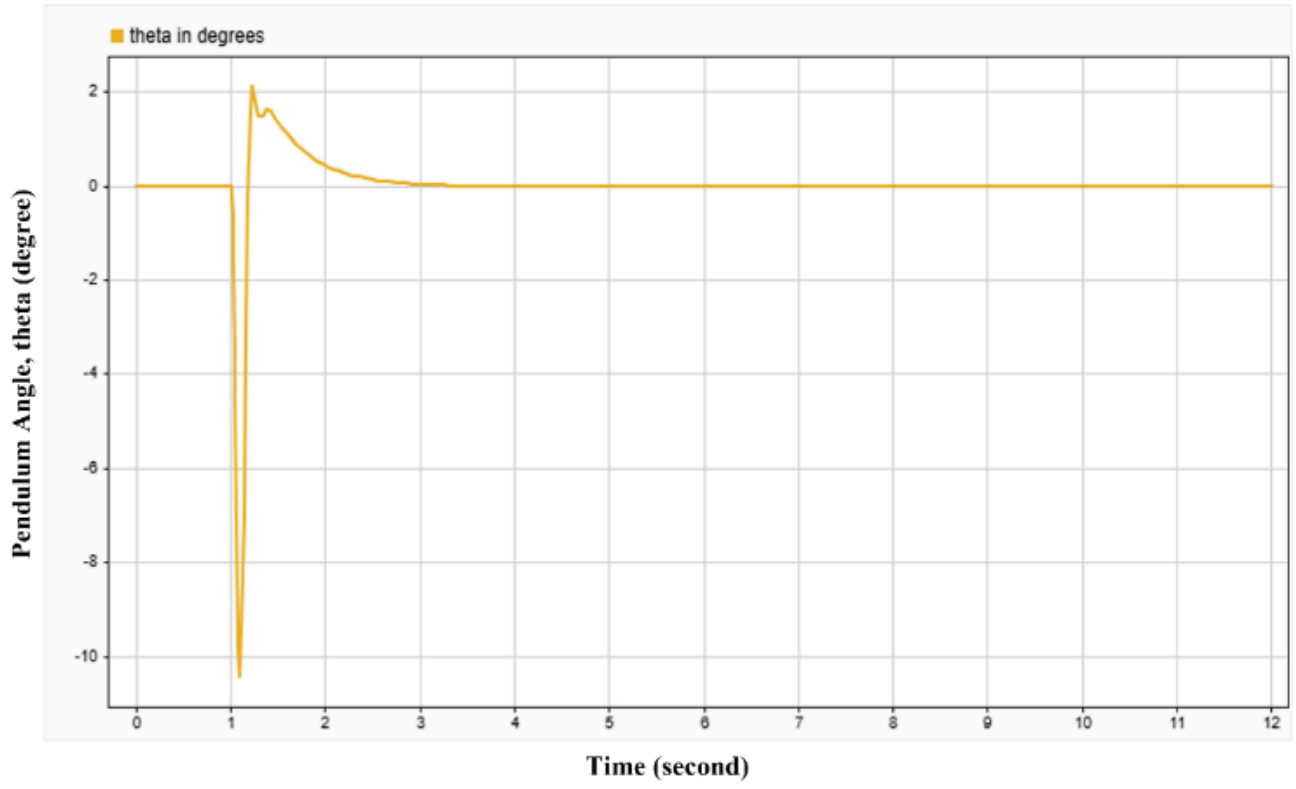

Fig. 5. Angle controller response for cart position input in Simulink.

Fig. 6 represents the corresponding position controller response. The cart position exhibited slower transient behavior, requiring approximately 7.3 seconds to settle within ±2% of the reference. The relatively long settling time reflects the tuning of the outer loop, which prioritizes stability and robustness over aggressive convergence. The initial backward motion to approximately 3 cm is very small compared to the rest of the data. The overshoot in this case was calculated to be approximately 4.71%.

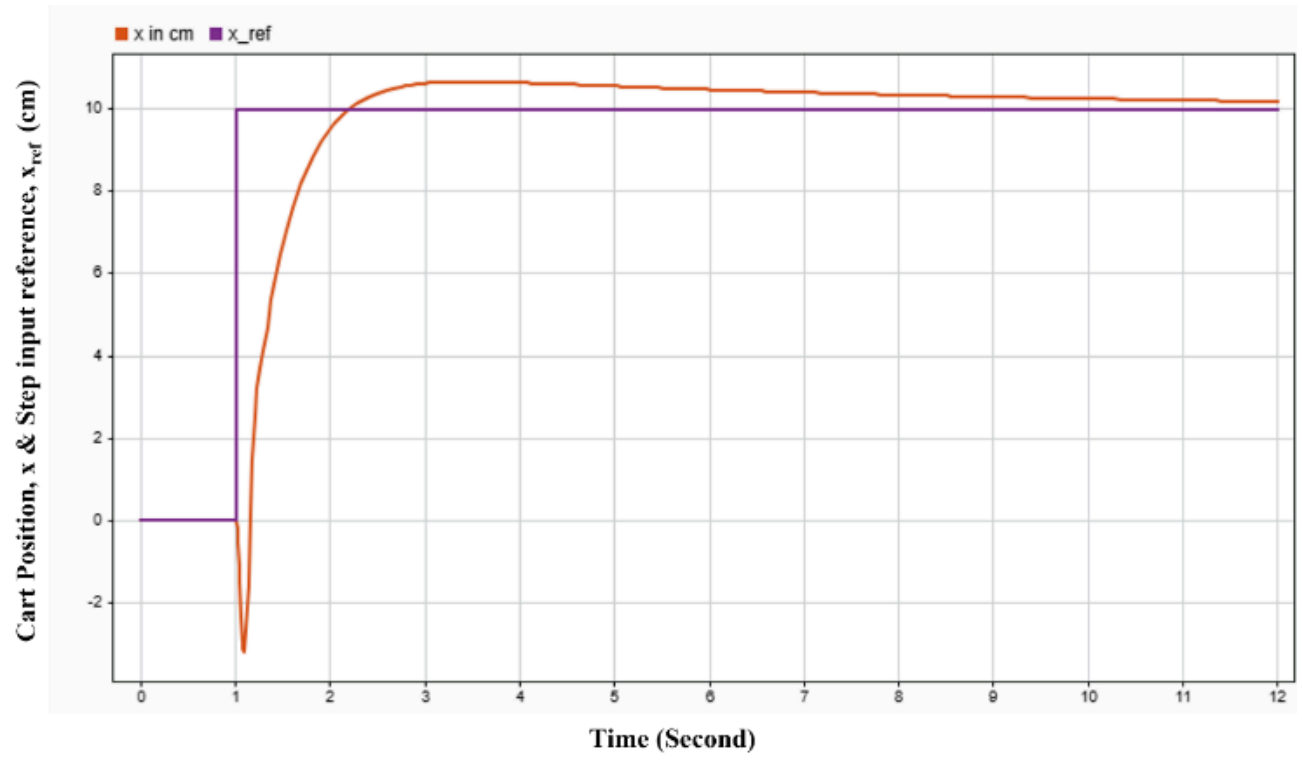

Fig. 6. Position controller response for the reference input of 10 cm.

Overall, the simulation results demonstrate that the cascade PID structure can stabilize the nonlinear inverted pendulum system and achieve convergence under idealized modelling assumptions. The force requirements of the plant for this control system are presented in Fig. 7, which further explains the reason behind the initial overshoot in both cases. As these forces remained within the system limits, it was considered acceptable.

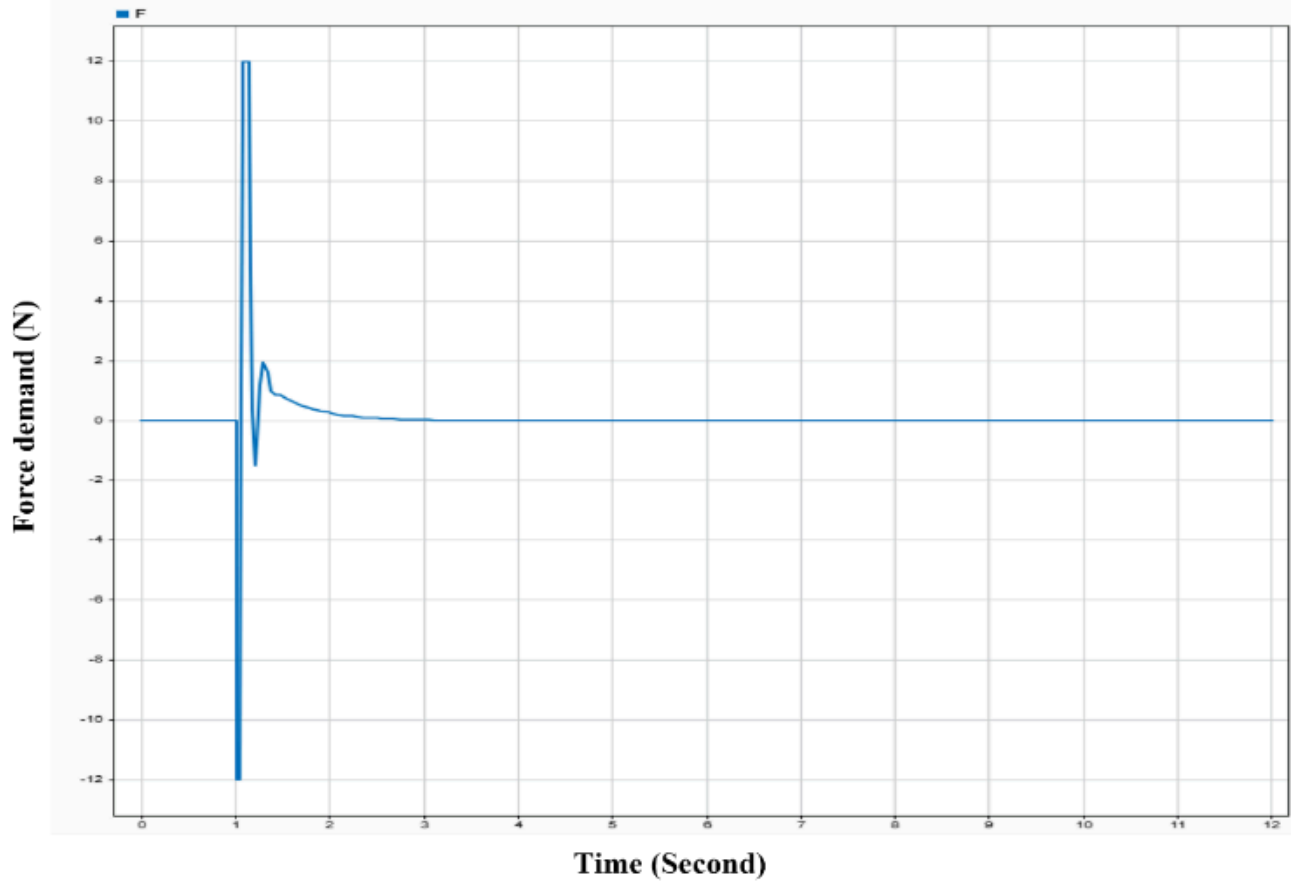

Fig. 7. Force demand of the angle controller.

### *B. Experimental Results*

Due to the large volume of data generated by the Arduino serial monitor, only a short segment of the recorded experimental data is presented for graphical representation. Real-time data filtering was not feasible, as the full data stream was required for controller operation. Before controller activation, the pendulum was manually positioned near the upright equilibrium and subsequently released. During this initialization phase, the system was subjected to small external disturbances (approximately 1 N).

Fig. 8 and Fig. 9 show representative experimental responses of both controllers. The pendulum angle was recorded in pulses, while the cart position was measured in centimetres. The pendulum angle response demonstrated a rapid reduction of angular deviations and remained in steady state throughout the experiment. From Fig. 8, the initial transient in the measured response reflects the manual positioning of the pendulum near the upright equilibrium before controller activation. A small peak value (approximately 8°) can be observed immediately after the pendulum was released, followed by convergence toward the upright position as the controller stabilizes the system.

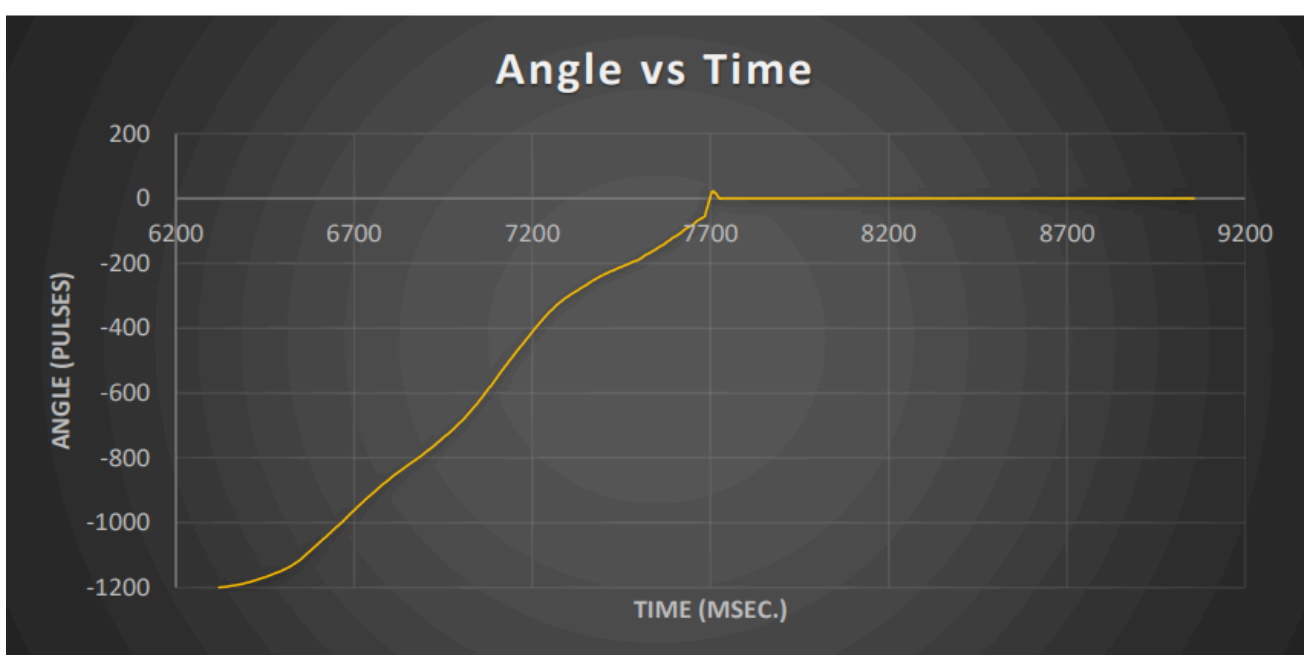

Fig. 8. Pendulum angle response (pulses) with respect to time (milliseconds).

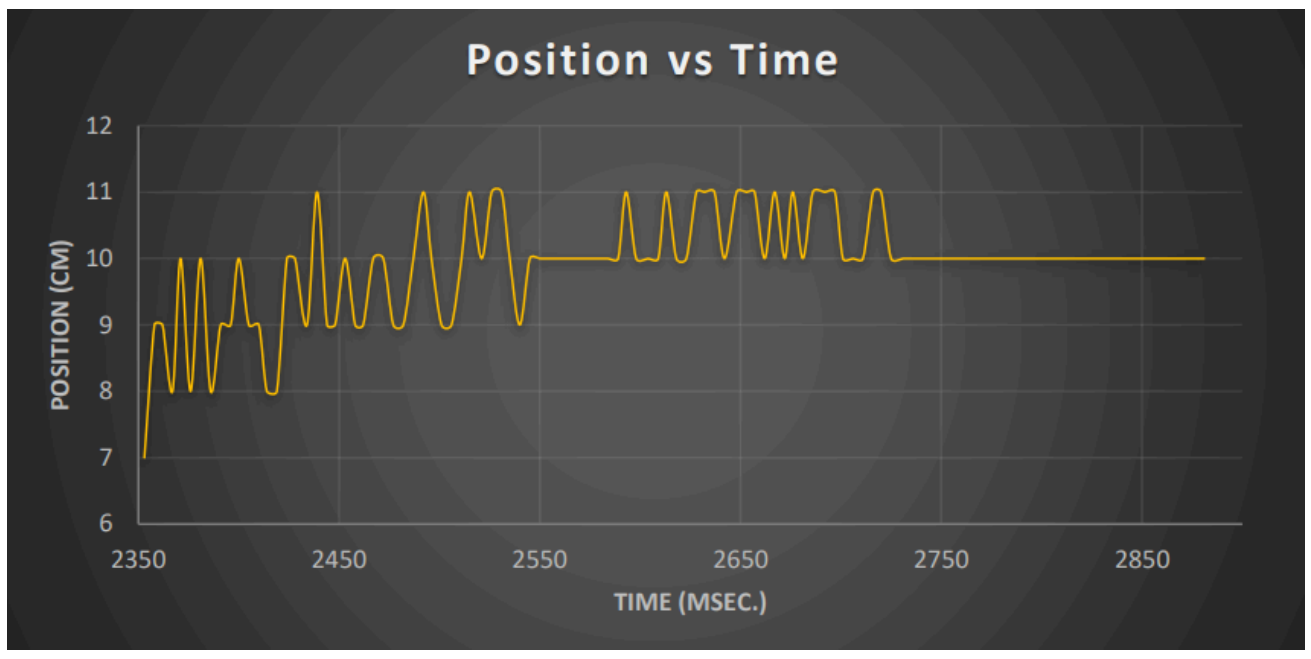

Fig. 9. Cart position tracking (cm) with respect to time (milliseconds).

Fig. 9 was obtained after the pendulum angle had settled near the upright equilibrium. At this time, small disturbances in the opposite direction of cart motion were applied very carefully so that they did not cause the angle to be changed more than approximately 10°. These disturbances caused the cart to move around the setpoint. Due to practical implementation limitations, simultaneous acquisition of pendulum angle and cart position data was not performed. This limitation was addressed by recording the angle controller response first and performing a separate experiment for the cart. In the second experiment, the cart was initialized at 7 cm and commanded to move to 10 cm while the pendulum was maintained near the upright equilibrium and then released. The cart position response exhibits a rapid transient decay before reaching steady-state. While the limited experimental window does not allow direct evaluation of steady-state behavior, the observed response confirms that the controller was able to track the input. The gains of both controllers are shown in Table III.

TABLE III. CONTROLLER GAINS

| Controller Gains | Simulation (Rounded-off values) | | Experiment (Exact values) | |
|---|---|---|---|---|
| | *Angle Controller* | *Position Controller* | *Angle Controller* | *Position Controller* |
| $K_p$ | 204.26 | −0.03 | 30 | 0.1945 |
| $K_i$ | - | - | 28.6 | 0 |
| $K_d$ | 3.70 | −0.24 | 0.1 | 0.000357 |
| Filter coefficient | 75.30 | 9.48 | - | - |

## C. Discussion

From Table III, it is evident that there is a noticeable difference between the simulated and experimental gains of the angle controller. This discrepancy can be attributed to four primary factors. First, unlike Simulink, the built-in PID library used in the experimental implementation does not provide an explicit filter coefficient option. This necessitated controller tuning through trial and error procedure. Second, control forces were computed directly in N and applied to the system in simulation. Whereas in the experimental setup, the controller output was implemented as a PWM command to the motor driver. This difference in actuation representation results in a mismatch between the effective control gains required in simulation and experiment. Third, the simulation employed exact measurements of both angular and linear positions, while the experimental system was subject to sensor noise and vibrations, leading to differences in the resulting error signals. Finally, frictional effects among the experimental cart, transport rail, and timing belt were not modelled in the simulation. The unmodeled friction also contributed to the shorter settling time observed in the experimental response.

The position controller gains are negative in the simulation. This explains the initial backward motion of the cart observed in the simulation before it corrects and converges to the reference position. On the other hand, negative gains could not be entered in the experimental setup as the built-in PID library only accepts positive PID gains. Therefore, the repeated disturbances applied in the opposite direction of motion acted as a replacement for the reverse motion required until the pendulum reached equilibrium. For small position references, the cascade PID controller exhibited satisfactory performance in simulation. However, for larger reference inputs, significant cart overshoot was observed. In such cases, the peak cart displacement increased substantially relative to the pendulum angle response, reflecting the increasing influence of nonlinear dynamics. This behavior also illustrates the inherent priority of the cascade control structure, in which the inner loop is strongly emphasized to maintain angle stability, even at the expense of large cart displacement. This effect is illustrated in Fig. 10, which also points out that the position command needs to be restricted to 25 cm for this system due to the limited track length.

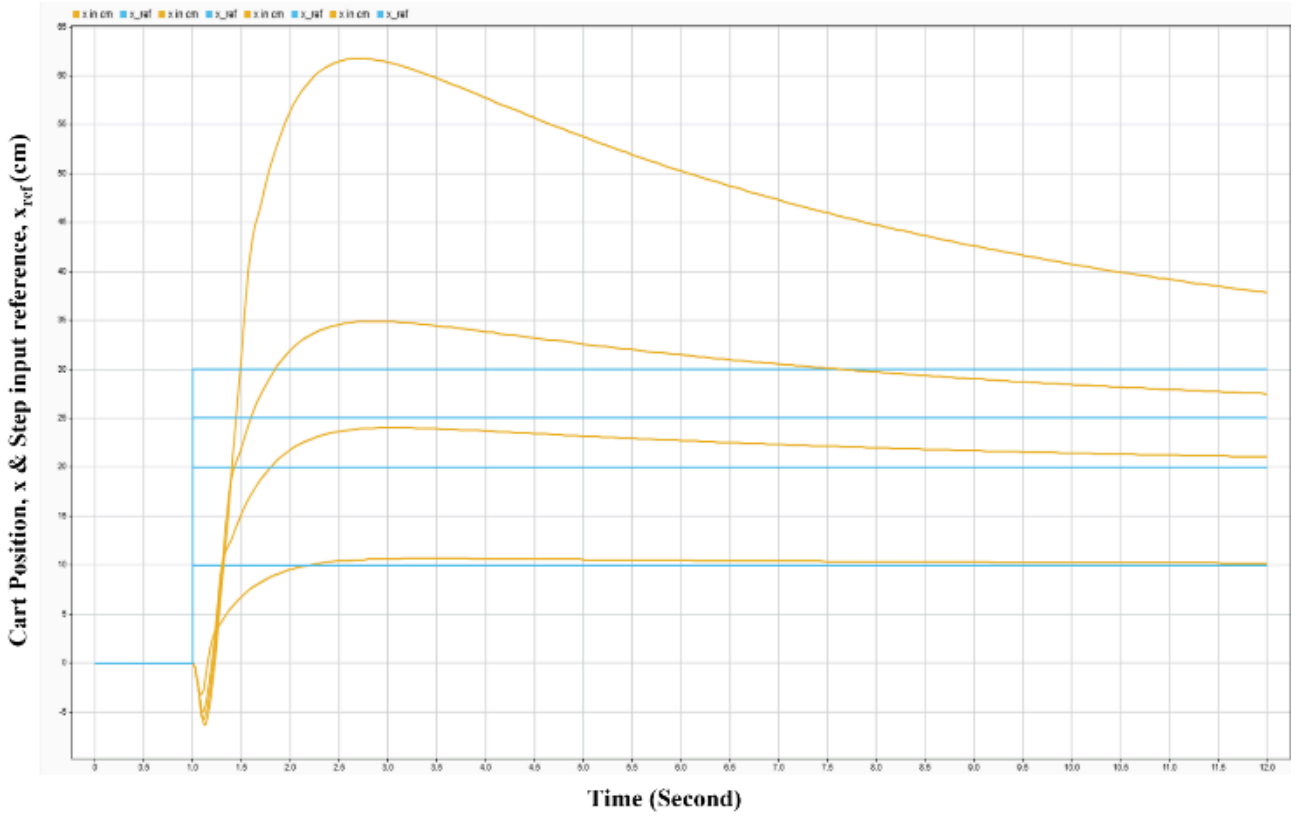

Fig. 10. Position tracking for 10, 20, 25, and 30 cm.

It was observed in experiment and simulation that the system could not reject disturbances. As the tuning did not include the effects of disturbances, this was to be expected. However, further tuning of the system did not result in any change, as presented in Fig. 11.

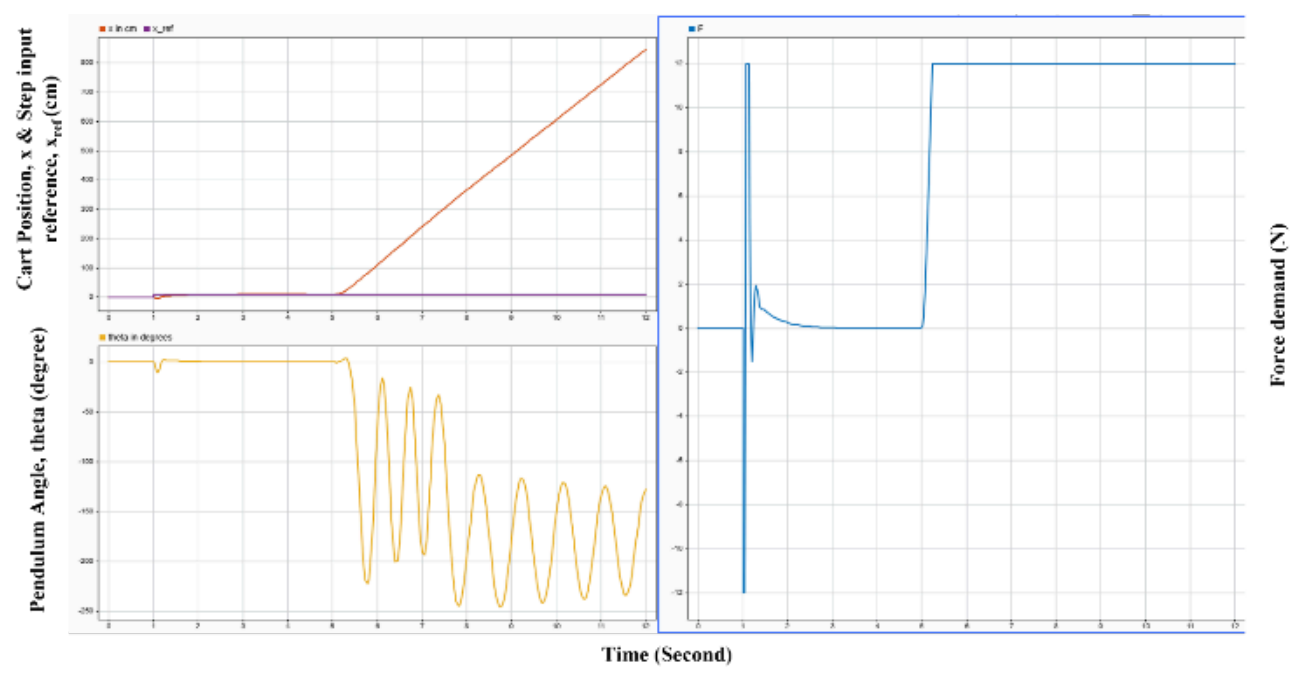

Fig. 11. Disturbance rejection failure with low pendulum mass (0.0374 kg).

Investigations into the system via simulation revealed that the pendulum mass was too small for this task. Increasing the pendulum mass to 0.6 kg and the length to 0.5 m produced the response shown in Fig. 12. The simulated response was obtained by applying a disturbance force of 1 N from the 6th second to the 7th second. Following this disturbance, a large transient cart displacement was observed while the pendulum quickly recovered steady-state. Thus, the controller prioritized pendulum stabilization over cart position regulation. Although the disturbance resulted in a significant cart displacement, the closed-loop system remained stable, and the cart position gradually returned to the reference. The system was able to stabilize with previous gains, but this also meant track length would have to be increased to 160 cm. This identified the limitations of cascade PID for the designed system.

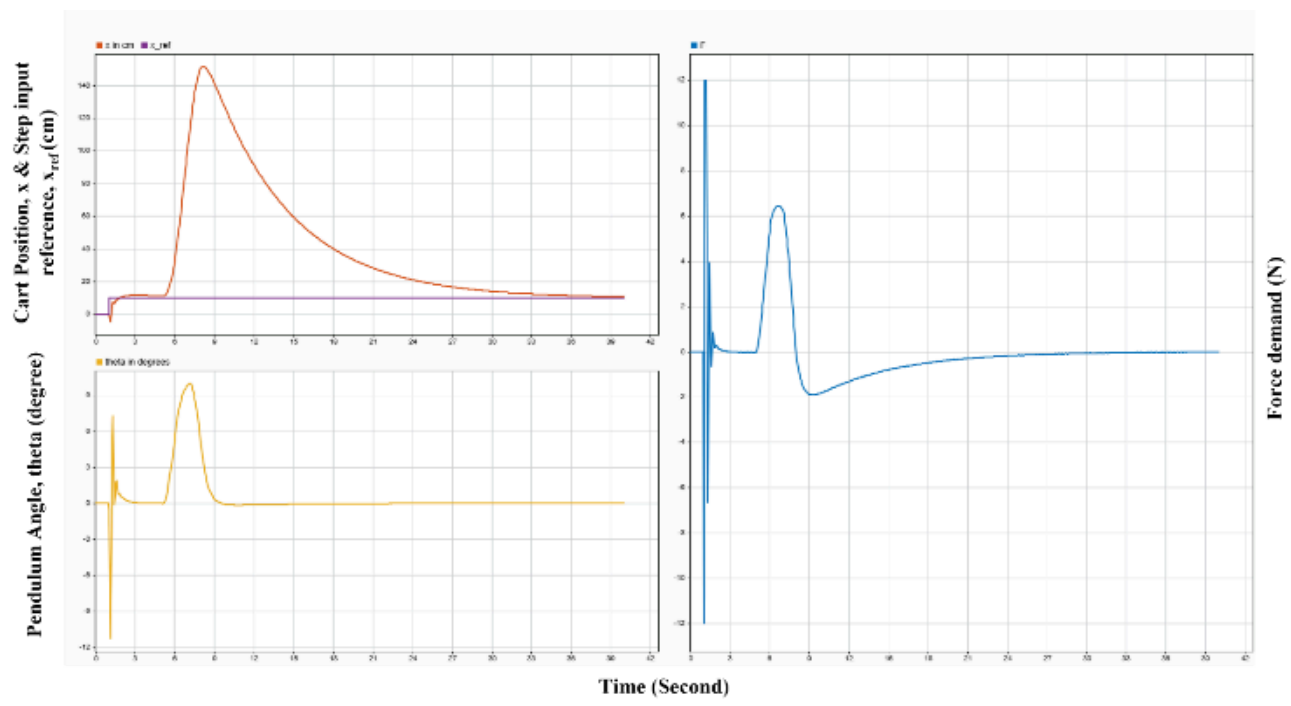

Fig. 12. Disturbance rejection of the system with increased pendulum mass and length by cascade PID.

To compare the performance of cascade PID with another method, the position controller was kept as PID with the same gains as before, while the angle PID controller was replaced with an LQR penalty of the form:

$$\int_0^{\infty}\left(2\theta^2(t) + x^2(t) + 0.001F^2(t)\right)$$

This resulted in the maximum overshoot in position being decreased to approximately 125 cm and a comparatively shorter settling time of approximately 20 seconds, as shown in Fig. 13. This structure also improved the initial high overshoot in angle and decreased force demand. While cascade PID can successfully reject disturbances applied to the system, it requires a higher track length.

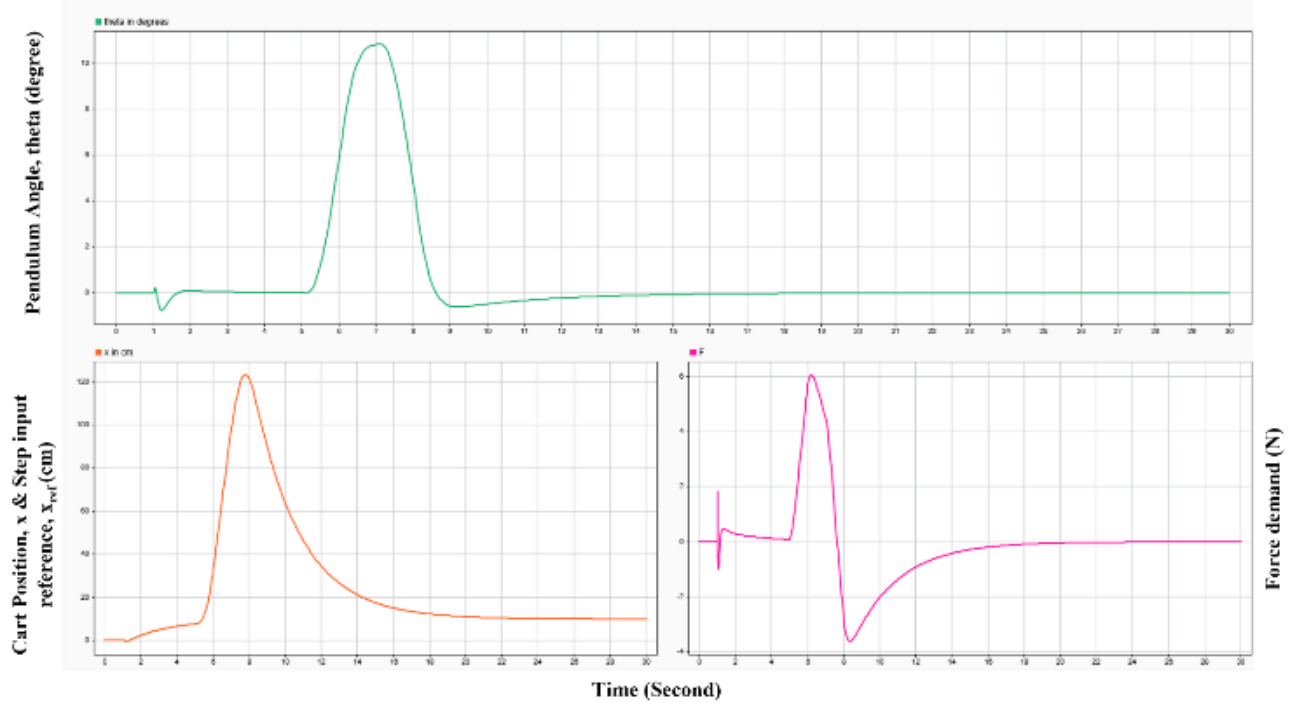

Fig. 13. Disturbance rejection with PID and LQR.

## VI. Conclusions

In this study, the cascade PID control method was implemented on the inverted pendulum on a cart through both simulation and experimentation. Simulation results showed that the system could be effectively stabilized. However, the pendulum mass and length needed to be increased from 0.0374 kg and 0.38 m to 0.6 kg and 0.5 m to achieve successful disturbance rejection. It was also found that stabilizing the system under disturbances with cascaded PID requires a relatively long track length of approximately 160 cm. Experimental results confirmed the applicability of the control structure, despite some discrepancies between the simulated and experimental gain values. The study discusses the causes of these differences (the absence of a filter coefficient in hardware, actuation representation differences, sensor noise, vibrations, and unmodeled frictional effects) and their impact on system performance. Additionally, it examines the effects of high demands on the inner loop controller in systems where a large outer loop overshoot is physically constrained. Replacing the inner loop controller with an LQR improved overall system performance and reduced position overshoot to 125 cm. Future work may focus on controlling the cart's acceleration and the pendulum's angular rate separately using nested PID loops to enhance system performance further.